\documentclass[12pt,a4paper,reqno]{article}
\usepackage{jheppub}
\usepackage{amssymb,eucal,xcolor,graphicx} 
\usepackage[USenglish]{babel}
\usepackage[T1]{fontenc}
\definecolor{refkey}{rgb}{0,0,1}
\definecolor{labelkey}{rgb}{1,0,0}
\usepackage{times,eulervm}
\usepackage{comment}
\allowdisplaybreaks[4]
\newcommand{\arxiv}{arXiv:}

\newcommand{\C}{\mathbb{C}}
\newcommand{\Q}{\mathbb{H}}

\newcommand{\id}{\mathsf{id}}

\newcommand{\D}{{D\mkern-13.5mu/\,}} 
\newcommand{\sptr}{{spectral triple}}
\newcommand*{\bfrac}[2]{\genfrac{}{}{0pt}{}{#1}{#2}}

\begin{document}
	\makeatletter
	\renewcommand{\@fpheader}{}
	\title{Fermion masses, mass-mixing\\ 
		and the almost commutative geometry\\
		of the Standard Model.}
	
	\author[a,1]{Ludwik D{\k a}browski,}
	\affiliation[a]{Scuola Internazionale Superiore di Studi Avanzati (SISSA), via Bonomea 265, \\ I-34136 Trieste, Italy.}
	\note{Partially supported through H2020-MSCA-RISE-2015-691246-QUANTUM DYNAMICS and through Polish support grant for the international cooperation project 3542/H2020/2016/2 and 328941/PnH/2016.}
	\emailAdd{dabrow@sissa.it}
	\author[b,c,2]{Andrzej Sitarz}
	\note{Supported by NCN grant OPUS 2016/21/B/ST1/02438}
	\affiliation[b]{Instytut Fizyki Uniwersytetu Jagiello{\'n}skiego, Stanis{\l}awa {\L}ojasiewicza 11, \\ 30-348 Krak{\'o}w, Poland.}
	\affiliation[c]{Institute of Mathematics of the Polish Academy of Sciences,
		\'Sniadeckich 8, \\ 00-656 Warszawa, Poland.}
	
	\emailAdd{andrzej.sitarz@uj.edu.pl}
	
	
	
	\abstract{
		We investigate whether the Standard Model, within the accuracy of current
		experimental measurements, satisfies the regularity in the form of Hodge 
		duality condition introduced and studied in \cite{DDS18}. We show that 
		the neutrino and quark mass-mixing  and the difference of fermion masses are necessary for this property. We demonstrate that the current data supports this new geometric feature of the Standard Model, Hodge duality, provided that all neutrinos are massive.
	}
	\maketitle
\section{Introduction}
The Standard Model of fundamental particles and their interactions
so successfully and accurately incorporates the vast amount of experimental data, that many theoretical efforts are nowadays redirected to proceed beyond it.
On the theoretical level it is a model of gauge fields (bosons) minimally coupled to matter fields (fermions) plus the Higgs field (boson). In a more mathematical terminology it can be described as a connection (multiplet of vectors) on (a multiplet of) spinors, plus a doublet of scalars. Of course, this layout necessitates the second quantization with gauge fixing, spontaneous symmetry breaking, regularization and perturbative renormalization.

Successful as it is, it is however inadequate for explaining (though somewhat constrains it) the contents of particles (especially 3 families) and the presence of several parameters. It does not include either the fourth known interaction: gravitation, together with its fundamental symmetry (general relativity).

There have been various attempts to settle some of the above issues, e.g. GUT based on a simple group $SU(5)$ or $SO(10)$, modern versions of Kaluza-Klein model with 'compactified' internal dimensions, and of course string theory and its further variants whose one of the targets remains the recovery of the Standard Model.

Another attempt is a minimal noncommutative description of the Standard Model (which we call for brevity {${\nu}$SM}). It has been formulated in the framework of noncommutative geometry by A.\,Connes\,et.al. \cite{Con95,Con96} and rather than {\em groups}  it is primarily based on {\em algebras}. It enriches the Gelfand-Naimark equivalence between topological spaces and commutative $C^*$-algebras,
and the Serre-Swan equivalence between vector bundles and modules. Namely also 
{\em smoothness}, {\em dimension}, {\em calculus}  and {\em metric} structure are encoded in algebraic terms using a spectral triple $(A, H, D)$ that consists of a $*$-algebra $A$  of operators on a Hilbert space $H$ together with an additional Dirac-type operator $D=D^\dagger$ on $H$. In addition,  an anti-unitary conjugation 
$J$ on $H$ is assumed, such that for any $a$ in $A$ the operator $JaJ^{-1}$ belongs to the commutant of $A$.

With an appropriately constructed geometry of such noncommutative type 
and the appropriate tools \cite{CM08} one finds then that 
$${\cal G}:=\{U=uJuJ^{-1}\,|\, u\in A, det\,U=1 \}  \simeq 
U(1)\!\times\! SU(2)\!\times\! SU(3)$$
yields the Standard Model gauge group (broken to $U(1)_{em}\times SU(3)$), with respect to which all the fundamental fermions in $H$ have the correct charges. Furthermore the 1-forms constructed from $a[D,b],\, a,b\in A$ and $J$ provide the Standard Model gauge fields $A_\mu, W^\pm, Z, G_\mu$ (from the part $\D$ of $D$),
plus the Higgs complex scalar (weak doublet) Higgs field
(from the part $D_F$ of $D$).

The merits of this formalism are that the gauge and Higgs field both arise as parts of a connection, that
it explains why only the fundamental representations of ${\cal G}$ occur, and that a simple spectral action $\hbox{Tr}\,f(D/\Lambda)$ reproduces the bosonic part of the Standard Model Lagrangian ${\cal L}_{SM}$ as the lowest terms of asymptotic expansion in $\Lambda$ and $\langle\phi, D\phi \rangle$ reproduces the 
(Wick-rotated) fermionic part, and moreover it naturally couples to gravity on $M$. Furthermore, it claims to predict a new relation among the parameters 
\cite{ChaCon} (see \cite{dlm14} for the Higgs mass estimates).

With the $\nu$SM matching so closely the Standard Model, it is clear that their immediate  predictive power should be comparable. On the conceptual level 
however $\nu$SM heralds quite an impressive message. Namely, its arena is the product 
of exterior (Wick rotated) space-time with a finite quantum  internal space. Though such a virtual space may not be directly observable, 
it reveals itself for instance due to internal component of the connection one forms, which can be identified with the Higgs field. 
Moreover, the Hilbert space $H$ built from the whole multiplet of fundamental matter fields (leptons and quarks), any of which is 
a Dirac spinor from the (Wick-rotated) space-time point of view, but with respect to the "flavour" degree of freedom it can be regarded 
as a field depending on the internal finite quantum direction. 

The goal of the present paper is to uncover what is the geometric nature of this internal part $H_F$ of the full Hilbert space $H$ in the Euclidean model
(see \cite{DDK16} and \cite{BS18} for the Lorentzian approach).
In particular we want to answer the question what type of fields on the quantum internal space are its elements. Already in \cite{DD14} it has been shown that they are certainly not quantum analogue of Dirac spinors. In  \cite{DDS18} it has been shown instead at least for the case of one generation that they are rather quantum 
analogue of de Rham differential forms. The  main result of the present paper is that this is indeed the case for the fully fledged Standard Model with
 three generations of particles.

The content of the paper is as follows. In section 2 we review the basic notions of the noncommutative geometry, define in this language the notion of Hodge duality, and introduce the notation for finite noncommutative geometries. In section 3 we present the discrete geometry in the description of the Standard Model. Section 4 is devoted 
to the analysis of the Hodge condition for the model introduced in the previous section, both in the 1 generation and in the 3 generation case. Finally, in section 5 we compare the
derived conditions with the experimental data and formulate some predictions based on the uncovered geometric structure.

\section{Preliminaries}

We start by recalling the notion of a basic definition of noncommutative geometry that extends the definition of a manifold and its two main classical models.

A spectral triple (noncommutative manifold), $(A, H, D) $,  consists of a $*$-algebra $A$ of operators on a Hilbert space $H$ and a Dirac-type operator $D=D^\dagger$ on $H$.  It satisfies certain analytic conditions which we will not dwell upon,
since they are automatically satisfied when the Hilbert space $H$ is finite dimensional, which will be of our particular interest here.
In addition  we assume an anti-unitary conjugation $J$ on $H$, such that 
for any $a$ in $A$ the operator $JaJ^{-1}$ belongs to the commutant of $A$.

A prototype, canonical example is a {\em spin} manifold $M$,
\begin{equation}\label{cst}
(C^\infty(M), L^2(S),\D ),
\end{equation}
where $C^\infty(M)$ is the algebra
of smooth complex functions on $M$, 
$L^2(S)$ are (square integrable) Dirac spinors on $M$,
and $\D$ is the usual Dirac operator on $M$.
The suitable $J_S$ is known as charge conjugation in physics.
Importantly these data can be characterized by $L^2(S)$ being the so called Morita equivalence $C^\infty(M)-\Gamma(\C l(M))$ bimodule. 
Here $\Gamma(\C l(M))$ is the algebra of Clifford fields,
generated by $C^\infty(M)$ and by the commutators $[\D,a]$, $a\in C^\infty(M)$, which are nothing but Clifford (or Dirac) multiplication by differential one-forms. The right action of $\alpha\in\Gamma(\C l(M))$ on $L^2(S)$  is given by $J_S\alpha J_S^{-1}$.
The Morita equivalence essentially means that these two algebras are in certain sense maximal one with respect to another on $L^2(S)$  
(and this is precisely so in case of finite dimensional $H$).
It is also worth to mention that \eqref{cst} fully encodes the geometric data on $M$, that can be indeed reconstructed \cite{Con13}.

Another natural \sptr\, is 
\begin{equation}
(C^\infty(M), L^2(\Omega(M)),d+d^* ),
\end{equation}
where\,$\Omega(M)$\,is\,the\,space\,of\,complex de\,Rham\,differential\,forms\,on\,
a closed oriented Riemannian manifold $M$,
$d$ is the exterior derivative and $d^*$ its adjoint 
with respect to the hermitian product induced by the metric $g$ on $M$.
The suitable $J_\Omega$ is the main anti-involution composed with complex conjugation.
Eminently these data can be characterized by $L^2(\Omega(M))$ being the so called Morita self equivalence $\Gamma(\C l(M))-\Gamma(\C l(M))$ bimodule. 
Here the left action comes, modulo the isomorphism $\Gamma(\C l(M))\approx \Omega(M)$ as vector spaces, from the left multiplication in  $\Gamma(\C l(M))$ 
and the right action is obtained from the left one by the similarity with $J_\Omega$. Furthermore $\Gamma(\C l(M))$ is again generated by $C^\infty(M)$ and by the commutators $[d+d^*,a]$, $a\in C^\infty(M)$.

\subsection{Quantum Clifford fields, spinors and forms}

Now quite as in \cite{LRV} also in the noncommutative context we regard the elements of the algebra $Cl_D(A)$ generated by $A$ and commutators $[D,A]$ as 'Clifford\,fields', since the elements $a\in A$ and $[D,a]$ play respectively the role of functions and differential one-forms on some 'quantum' (virtual) space.
Next, motivated by the two above classical cases we call a general (not necessarily commutative) \sptr\, $(A,H,D)$ with conjugation $J\,$  
\underline{\em spin} 
when $H$ is a {Morita} equivalence \cite{Ply86} $\C l_D(A)$-$A$ bimodule
and the right action of $a\in A$ is $Ja^*J^{-1}$ \cite{DD14}.
Furthermore, the elements of $H$ are called \emph{quantum Dirac spinors}. 
On the other hand, following \cite{DDS18},  we call $(A,H,D)$ with conjugation $J\,$
\underline{\em Hodge},
when $H$ is a {Morita} equivalence $\C l_D(A)$-$\C l_D(A)$ bimodule
and the right action of $\alpha\in \C l_D(A)$ given by $J\alpha^*J^{-1}$. 
Furthermore the elements of $H$ are called \emph{quantum de\,Rham forms}.

The physical fermion fields that represent the elementary particles are Dirac spinors whereas, as shown in \cite{DDS18}, the finite noncommutative geometry of the Standard Model bears resemblance to the Hodge type.

\subsection{Finite noncommutative geometries}

We present basic facts, notation and conventions about operators (matrices) 
on a finite-dimensional Hilbert space $H$ that are needed for our purposes. 

We denote by $M_j$ the algebra of complex $j\times j$ matrices; in particular $M_1=\C$.
Let $a_1,\ldots, a_k$ be matrices such that 
$a_i \in M_{n_i}$. By
\begin{equation}\label{rep}
a_1^{(p_1)} \oplus \cdots \oplus a_k^{(p_k)}
\end{equation}
we mean a block diagonal matrix in $M_N$, where
$N = n_1 p_1 + \cdots +n_k p_k$, where the first matrix $a_1$ appears
block-diagonally $p_1$ times, then $a_2$ appears $p_2$ times etc.
For the zero matrix we always assume that it acts on $\C$ and hence 
$0^{(k)}$ means $k\times k$ matrix of zeroes.

The matrices \eqref{rep} form the algebra 
$ A=M_{n_1}^{(p_1)} \oplus \cdots \oplus M_{n_k}^{(p_k)}$
which is an isomorphic copy (faithful representation on the $N$-dimensional Hilbert space $H$) 
of the algebra  
\begin{equation}\label{alg}
 M_{n_1} \oplus \cdots \oplus M_{n_k}
 \end{equation}
and to simplify the notation  
we will occasionally identify them.

The commutant of $A$, by which we understand the maximal 
subalgebra of $M_N$  that commutes with $A$ on $H$, is then
\begin{equation}
\label{comm}
A'= M_{p_1}^{(n_1)} \oplus \cdots \oplus M_{p_k}^{(n_k)}.
\end{equation}
This follows directly from the Schur's lemma applied to matrix algebras.
We as well call {\em commutant} of \eqref{alg} the isomorphic copy 
$M_{p_1}(\C) \oplus \cdots \oplus M_{p_k}(\C)$ of the algebra 
\eqref{comm}.
We shall also use representations that are equivalent
via a permutation of the basis of $H$, clearly, the commutant $A'$ of $A$ is insensitive to such 
operations up to an isomorphism. 

For a finite  spectral triple (see \cite{PS96} for details) $(A,H,D)$, i.e. 
with a finite dimensional Hilbert space $H$,  the Hodge duality condition of \cite{DDS18} in terms of Morita equivalence, as stated above, can be simply 
formulated as the duality between certain algebra and its commutant. 
Namely, let $\C l_D(A) $ be 
the algebra generated by $A$ and the commutators $[D,A]$.
We say that $(A,H,D)$ satisfies
the Hodge duality if the commutant $(\C l_D(A))'$ of $\C l_D(A)$ is anti-unitary equivalent to $\C l_D(A)$, i.e. there is a norm preserving antilinear operator $J$
on $H$ such that 
\begin{equation}
(\C l_D(A))' = J \C l_D(A) J^{-1}.
\end{equation}

For finite dimensional algebras this condition can be simplified a lot. First of all, both $(\C l_D(A))'$ and $J \C l_D(A) J^{-1}$ are in fact  finite direct sum of full matrix
algebras, which are represented on the same, finite-dimensional Hilbert space. 
Therefore, to check the Hodge duality it is sufficient to use the exact form of the representation and compute the commutant of a finite dimensional algebra using
the formula (\ref{comm}). 

We now pass to verify the Hodge condition for the noncommutative geometry
underlying the above formulation of the Standard Model with a choice of 
its algebra and Dirac operator.

\section{Finite Geometry of the Standard Model}

The ``almost commutative'' geometry \cite{vS15} of the Standard model is described by the product of the canonical spectral triple \eqref{cst} with 
with the 'internal' {\em finite} one
\begin{equation}
(A_F, H_F, D_F).
\end{equation}
We consider the case with Dirac neutrinos and with no leptoquarks, that is with separate masses and mixing matrices for leptons and for quarks. 

The Hilbert space that describes the matter fields is 
\begin{center}
$L^2(S)$
$\otimes\,  H_F,$ 
 \end{center}
where
$$
H_F= \C^{96} =: H_f\otimes \C^{3},
$$
with $\C^{g}$ corresponding to $g$ generations ($g=3$ as currently observed), and 
$$H_f= \C^{32}\simeq M_{8\times 4}(\C)$$
with basis labeled by particles and antiparticles, 
we arrange in the following way
\begin{equation}
\label{reps}
\begin{bmatrix}
\nu_R & u^1_R & u^2_R & u^3_R \\
e_R   & d^1_R & d^2_R & d^3_R \\
\nu_L & u^1_L & u^2_L & u^3_L \\
e_L   & d^1_L & d^2_L & d^3_L\\
\hspace{1pt}
\bar\nu_R & \bar e_R & \bar\nu_L & \bar e_L \\
\bar u^1_R & \bar{d}^{\,1}_R & \bar u^1_L & \bar{d}^{\,1}_L \\
\bar u^2_R & \bar{d}^{\,2}_R & \bar u^2_L & \bar{d}^{\,2}_L \\
\bar u^3_R & \bar{d}^{\,3}_R & \bar u^3_L & \bar{d}^{\,3}_L
\end{bmatrix} 
\end{equation}
where 1,2,3 are the {\em color} labels. For convenience, some other arrangements will also be used. Thus from the (Wick-rotated) space-time point of view,
the matter field are Dirac spinors, while the full "flavour" multiplet of them taken all together can be  thought of as a field with internal degrees of freedom 
on some finite quantum (virtual) space $F$.\footnote{We refer to \cite{DDK16} for the treatment of the apparent doubling due to the external and 
internal chiralities and antiparticles.}

More precisely  the underlying arena of $\nu$SM is
described "dually" by the algebra $C^\infty(M)\otimes A_F$, 
where $A_F$, isomorphic to $\C\oplus\Q\oplus {M_3(\C)}$, 
is realised diagonal in generations and acts on $H_f$ as 
the left multiplication of the columns of the matrix \eqref{reps} by the matrices:
\begin{equation}\label{eq:8t8}
\left[\begin{matrix}
\left[\!
\begin{array}{c|c}
\begin{matrix} \,\lambda\; & 0\;  \\ 0 & \bar\lambda \end{matrix} &
\begin{matrix} \;0\; & 0\;  \\ 0 & 0 \end{matrix} \\
\hline
\begin{matrix} \;0\; & \;0\; \\ 0 & 0 \end{matrix} & q
\end{array}
\!\right] \!\! & {\huge 0_4} \\ 0_4 & \!\!
\left[\!
\begin{array}{c|ccc}
\lambda & 0\; & 0\; & 0\; \\
\hline
\begin{matrix} \;0\; \\ 0 \\ 0 \end{matrix} && m
\end{array}
\!\right]
\end{matrix}\right]\; ,
\end{equation}
where $\lambda \in  \C$, a quaternion $q$ is written as a $2\times 2$ complex matrix, and $m\in {M_3}$. 

Note that $A_F$ is a real $*$-algebra; however, we shall work with its complexification $A_F^\C$  given by \eqref{eq:8t8}
where  $\bar\lambda$ is replaced by an independent $\lambda'\in \C$,
and the quaternion $q$ is replaced by a complex matrix in $M_2$.

The real conjugation is $J=J_S\otimes J_F$, where $J_F$  on $H_f$ is
\begin{equation}\label{eq:JF}
J_F\begin{bmatrix} v_1 \\ v_2 \end{bmatrix}=\begin{bmatrix} v_2^* \\ v_1^* \end{bmatrix} \ .
\end{equation}

Finally, the Dirac operator is\,  $D=\D\otimes\id +\gamma_S\otimes D_F$, where 
we can consider only the part of $D_F$ that does not commute with the algebra.

A convenient way to write both the action of the (suitably complexified) internal algebra of the Standard Model as well as the Dirac operator follows 
directly from \eqref{eq:8t8}:
\begin{equation}\label{eq:8t82}
A_F =  (M_1 \oplus M_1 \oplus M_2)^{(4g)} 
\oplus 
(M_1 \oplus M_3)^{(4g)},
\end{equation}
with the first and last $M_1$ summands identified,
i.e. the element
$z \oplus w \oplus h \oplus m \in  M_1 \oplus M_1 \oplus M_2 \oplus M_3$
is acting as 
$$ \left( z \oplus w \oplus h \right)^{(4g)}
\oplus (z \oplus m)^{(4g)}. $$
Note that this algebra is, in fact, equivalent to $M_1^{(8g)} \oplus M_1^{(4g)} \oplus M_2^{(4g)} \oplus M_3^{(4g)}$ but it is convenient for our purposes to permute some of the subspaces in the Hilbert space.

Analogously the (relevant part) of the Dirac operator $D_F$ 
can be written equivalently as
\begin{equation}\label{tdf2}
\tilde D_F = \left( D_{l} \oplus D_{q}^{(3)} \right) \oplus 0^{(16 g)}\ ,
\end{equation}
where where $D_l, D_q \in M_{4g}$ are positive mass matrices for leptons and 
quarks, respectively. We assume them to be of the form
\begin{equation}\label{DlDq}
D_{l} = 
\left( 
\begin{array}{c|c}
\begin{matrix} 0 \;& \; 0 \\ 0 \;& \;0 \end{matrix} &
\begin{matrix} \Upsilon_\nu & 0 \\ 0 & \Upsilon_e \end{matrix} \\
\hline
\begin{matrix}  \Upsilon_\nu^* &  0  \\ 0 & \Upsilon_e^* \end{matrix} &
\begin{matrix} \;0\; & \;0\; \\ 0 & 0 \end{matrix} 
\end{array}
\!\right),
\quad
D_{q}= 
\left( 
\begin{array}{c|c}
\begin{matrix} 0 \;& \; 0 \\ 0 \;& \;0 \end{matrix} &
\begin{matrix} \Upsilon_u & 0 \\ 0 & \Upsilon_d\end{matrix} \\
\hline
\begin{matrix}  \Upsilon_u^* &  0  \\ 0 & \Upsilon_d* \end{matrix} &
\begin{matrix} \;0\; & \;0\; \\ 0 & 0 \end{matrix} 
\end{array}
\!\right),
\end{equation}
with various $\Upsilon$'s\;$\in M_g$, where $g$ is the number of
generations (experimental data attest that $g=3$).  We assume 
$\Upsilon$'s to be unitarily similar to diagonal matrices. 

\section{Hodge condition}

Concerning the matrix $D_F$ that plays the role of the internal Dirac operator there are various possible choices, however, not all of them will result in the Hodge duality.  We start with the simpler case $g=1$ in order to recast the results 
of \cite{DDS18} in our present notation and conventions. Next we will pass to the physically relevant case $g=3$. 

Before we start let us note that out of the various possibilities in 
\cite{DDS18} which resulted in the models that satisfied Hodge duality 
we focus on one, physically relevant with the $D_F$ given by 
\eqref{DlDq}. As it has been already demonstrated in \cite{DDS18} that for such Dirac operator 
the so-called second order condition \cite{FB14a} is satisfied, we know that $(\C l_D(A))'$ contains $J\,\C l_D(A) J^{-1}$. 
Therefore, the problem of verification whether these two algebras are equal can be easily reduced to simple computation of dimensions, using the arguments from  section 2. 
Since all the algebras are finite-dimensional matrix algebras
represented on the same Hilbert space it is sufficient to compute  the
algebra $(\C l_D(A))'$ and compare it with $(\C l_D(A))$.

\subsection{One generation}

In this case the various $\Upsilon$'s in \eqref{DlDq} are just complex numbers 
and so the $D_{l}$ and $D_q$ matrices are just $4\times 4$ complex matrices 
acting on the spaces of leptons and quarks, where $\Upsilon_e$ is the 
electron mass and $\Upsilon_\nu$ is the neutrino mass (and similarly 
for $D_{q}$).

Since ${\C l_D(A)}$ contains $M_1^{(4)} \oplus M_3^{(4)}$ 
by \eqref{comm} the commutant ${\C l_D(A)}'$ of ${\C l_D(A)}$ must contain 
$M_4 \oplus M_4^{(3)}$, and, if the Hodge duality is satisfied, so must 
$\C l_D(A)$. However, $\C l_D(A)$ contains $M_1^{(4)} \oplus M_3^{(4)}$ and
two algebras generated respectively by $M_1 \oplus M_1 \oplus M_2$ and
$D_l$, and $(M_1 \oplus M_1 \oplus M_2)^{(3)}$ and $D_q^{(3)}$.
Thus the only possibility that the Hodge condition holds is when 
these two algebras are $M_4$ and $M_4^{(3)}$, respectively. 
This happens only when independently all $z \oplus w \oplus h 
\in M_1 \oplus M_1 \oplus M_2$ and $D_l$, as well as $z \oplus w \oplus h$ 
and $D_q$, each generate $M_4$. It is easy to notice that sufficient 
and necessary condition for this is that all four masses $\Upsilon$'s 
are different from zero.

Moreover, to guarantee that the algebra generated by 
$(M_1 \oplus M_1 \oplus M_2)^{(4)}$ and $D_l \oplus D_q^{(3)}$
is indeed $M_4 \oplus M_4^{(3)}$ one needs to impose certain conditions 
that relate the matrices $D_l$ and $D_q$, which enforces some further 
restrictions on the masses $\Upsilon$'s. 
Namely,  there can not be any nontrivial
matrix in $ M_{8}$ that commutes both with the algebra 
$(M_1 \oplus M_1 \oplus M_2)^{(2)}$
and the operator $D_l\oplus D_q$. Since without loss of generality 
it can be taken hermitian (as $D_l$ and $D_q$ are hermitian), then
it must thus have the form  
$$ \left( \begin{array}{cc} c_1 1_{4} &\, Q \\ Q^* &\, c_2 1_{4} \end{array} \right),$$ 
where $Q$ is is non-zero matrix in $(M_1 \oplus M_1 \oplus M_2)$, 
$c_1,c_2$ are complex numbers and 
$$Q= Q_1 \oplus Q_2 \oplus (1_2 \otimes Q_3) $$ 
with each $Q_1, Q_2, Q_3 \in \C$. It is not difficult to see that 
the inequalities $|\Upsilon_\nu|\neq |\Upsilon_u|$  and 
$|\Upsilon_e|\neq |\Upsilon_d|$ are both sufficient and necessary to 
assure that the only solutions are $Q_1=Q_2=Q_3=0$, which reproduce  
the conditions of \cite{DDS18} for the Hodge property when $g=1$.
 
\subsection{Three generations}

In this section we shall generalize some of the results of \cite{DDS18} and the previous sections first to an arbitrary number $g$ of generations,
and then will concentrate on the case $g=3$.

The Hilbert space is now just the $g$-multiple of the Hilbert space considered above, or, what is the same, tensor with $\C^g$. In other words every matrix 
element becomes now a matrix in $M_g$. 

The algebra of the Standard Model acts in each case diagonally, that it's representation is just diagonally $g$-copies of the earlier considered representations. 

As mentioned we consider the case with Dirac neutrinos and with no leptoquarks
\cite{PSS99}, that is with separate masses and mixing matrices for leptons and for quarks and thus $A_F$ is given by \eqref{eq:8t82} and $\tilde D_F$ by \eqref{tdf2}.

Observe that most of the arguments that were used in the previous section can be easily adapted to our case. So, the commutant of the algebra generated by $A_F$ and $\tilde D_F$ has certainly two copies of $M_{4g}$. 
Therefore, the only possibility that the generated algebra is Hodge selfdual 
is if the algebra generated by
$$  
(M_1 \oplus  M_1 \oplus M_2)^{(g)} {\,\, \rm and}\,\,  D_{l}\ , $$
and the algebra generated by
$$ 
(M_1 \oplus  M_1 \oplus M_2)^{(g)} {\,\, \rm and}\,\,  D_{q}\ , $$
are (isomorphic) to the full matrix algebras $M_{4g}$. Note that 
this are only partial conditions for the Hodge property
whereas we need later examine the condition that the 
algebra
$(M_1 \oplus  M_1 \oplus M_2)^{(2g)}$ (isomorphic with
$(M_1 \oplus  M_1 \oplus M_2)^{(g)}$) 
and $D_l\oplus D_{q}\ , $ generate the full matrix algebra $M_{4g}
\oplus M_{4g}$, which depends on possible relations between 
$D_{l}$ and $D_{q}$ and which can break Hodge duality.

We focus on the physically relevant case where the matrix $D_{l}$ (and 
similarly $D_{q}$) are of the form \eqref{DlDq}, when acting on the 
spaces of leptons and quarks, this time, however, with $\Upsilon$'s 
being hermitian mass matrices in $M_g$. 

\subsubsection{Partial conditions}

We start with leptons and use a simple argument to check whether the algebra generated by $A_{l}, D_{l}$ is a full matrix algebra. 
  
With this we may use next just Schur's lemma in the following form: the fact that
the algebra generated by some matrices is a full matrix algebra is equivalent 
to the fact that the only matrix that commutes with them is a multiple 
of identity.

A general matrix that commutes 
with $A_{l}$ has a form $P_1 \oplus P_2 \oplus \tilde{P}_3$,
where $P_1,P_2 \in M_g$ and $\tilde{P}_3 = 1 \otimes P_3 \in
 M_2 \otimes M_g$. If it commutes with $D_{l}$ then:
\begin{equation}
\begin{aligned}
&P_1 \Upsilon_\nu = \Upsilon_\nu P_3, \qquad 
&P_2 \Upsilon_e = \Upsilon_e P_3, \\
& P_3 \Upsilon_\nu^* = \Upsilon_\nu^* P_1, \qquad 
& P_3 \Upsilon_e^* = \Upsilon_e^* P_1. 
\end{aligned} 
\label{pcond}
\end{equation}
{}From these equations we immediately infer that $P_1$ and $P_3$ must commute with $\Upsilon_\nu \Upsilon_\nu^*$ (note that since both 
$\Upsilon$ matrices are unitarily similar to diagonal matrix then they are normal) whereas  $P_2$ and $P_3$ must commute with $\Upsilon_e \Upsilon_e^*$. 

If the two matrices $\Upsilon_\nu \Upsilon_\nu^*$ and  $\Upsilon_e \Upsilon_e^*$
generate the full matrix  algebra $M_g$ then by Schur's lemma the matrix $P_3$ must be proportional to identity matrix.  However, by looking on the form of 
equations \eqref{pcond} we see that if $\Upsilon_\nu$ is not invertible then
one can find $P_1$ that satisfies them, similar argument holds, of course, for
$\Upsilon_e$ and $P_2$. Therefore, only if both $\Upsilon_\nu$ and
$\Upsilon_e$ are invertible and  the pair $\Upsilon_\nu\Upsilon_\nu^*$ and  $\Upsilon_e \Upsilon_e^*$ generate the full matrix algebra $M_g$  it follows consequently that  $P_1$ and $P_2$ must be equal to $P_3$, and be proportional to the identity. 

If the only solution for $P_1 \oplus P_2 \oplus \tilde{P}_3$ is a matrix proportional 
to identity then the algebra generated by $A_{l}$ and $D_{l}$ is indeed a full matrix algebra. Observe that these conditions are truly independent as two matrices 
can generate a full matrix algebra even though they are not invertible.

Similar arguments will also hold for the quarks: it suffices (and is necessary) that the two matrices $\Upsilon_u \Upsilon_u^*$ and  $\Upsilon_d \Upsilon_d^*$ generate the full matrix algebra $M_g$ and that they are invertible to assure that the algebra generated by  $A_{q}, D_{q}$ is a full matrix algebra.

Out of the above condition, the invertibility of $\Upsilon$ is easy to verify, as
since they are assumed to normal then they are diagonalizable and the condition 
can be rephrased that neither of them have a zero eigenvalue. 

To verify the second condition we see that we have therefore reduced the 
problem to the case of two hermitian matrices and the question when they
generate a full matrix algebra.

Let us briefly remind when two hermitian matrices, $A,B$ in $M_3(\C)$ (as we are 
dealing with the physical case then $g=3$) generate a full matrix algebra. 
The sufficient and necessary condition, which directly follows from  a  
result obtained by Burnside in 1905 \cite{Burn} is that they do not 
share a common eigenvector (the theorem states that there is no 
common invariant subspace but since the matrices are hermitian if 
there exists an invariant subspace its complement is also invariant and 
hence there would necessarily exist an invariant subspace of dimension 
$1$).
	
Now, since the problem is invariant under the simultaneous adjoint action 
of $U(3)$, without loss of generality we can assume that we work 
in a basis in which one of the matrices, say $A$, is diagonal. Next, 
the matrix $B$ can be written in the form $U b U^*$, where $U\in U(3)$ 
and $b$ is also diagonal in the chosen basis.
The condition that both $A, B$ share a common eigenvector can be translated 
to the property that the matrix $U$ maps at least one of the basis vectors 
to another basis vector.  Indeed, let $e$ be one of the basis vectors, then by construction it is an eigenvector of $A$.  If $ U b U^* e = \lambda  e$ then 
taking $f$ such that  $e = U f$ we have $b f = \lambda f$.  However, since 
by assumption $b$ was diagonal then $f$  is again one of the 
basis vectors. 

If, we assume that all eigenvalues of $A$ are different from each other then we only need to check the matrix elements of $U$ in the chosen basis, in which
$A$ is diagonal. If no matrix element of $U$ is of modulus $1$ (while at the same time other matrix elements in the same row and in 
the same column are $0$) then $U$ does not map one of the basis vectors to another one. Equivalently, one
can reformulate the condition in the following way: no permutation of the basis leads to the block diagonal matrix 
of $U$ with rank of  the largest block strictly less than $3$.

\subsubsection{Full conditions}

Finally, we analyse when the algebra generated by 
$(M_1 \oplus M_1 \oplus M_2)^{(2)}$ 
(isomorphic with $(M_1 \oplus  M_1 \oplus M_2)^{(g)}$)
and the matrix $D_{l} \oplus D_{q}$
is exactly $M_{4g} \oplus M_{4g}$,
which imposes certain conditions that relate $D_{l}$ and $D_{q}$. We assume that both matrices are of the chosen form \eqref{DlDq}
and that each of the generates
a full matrix algebra together with $A_{l}$ and $A_{q}$, respectively.

If the algebra generated by $(M_1 \oplus M_1 \oplus M_2)^{(2)}$ and $D_{l} \oplus D_{q}$  is smaller than $M_{4g} \oplus M_{4g}$ then there 
must exist a matrix in $ M_{8g}$ that commutes both with the 
algebra $(M_1 \oplus M_1 \oplus M_2)^{(2)}$ and the operator 
$D_{l} \oplus D_{q}$, and which without loss of generality can be taken 
hermitian (as $D_l$ and $D_q$ are hermitian).
It must thus have the form  
$$ \left( \begin{array}{cc} c_1 1_{4g} &\, Q \\ Q^* &\, c_2 1_{4g} \end{array} \right),$$ 
where $Q$ is is non-zero matrix in $(M_1 \oplus M_1 \oplus M_2) \otimes M_g$, 
$c_1,c_2$ are complex numbers and 
$$Q= Q_1 \oplus Q_2 \oplus (1_2 \otimes Q_3) $$ 
with each $Q_1, Q_2, Q_3 \in M_g$. 

We obtain:
$$ D_l Q = Q D_q, \qquad D_q Q = Q D_l, $$
which leads to:
\begin{eqnarray*}\label{QQQ}
&\Upsilon_\nu    Q_3  =  Q_1 \Upsilon_u,  \qquad &
\Upsilon_e        Q_3  = Q_2 \Upsilon_d, \\
& \Upsilon_\nu^* Q_1  = Q_3 \Upsilon_u^*, \qquad &
\Upsilon_e^*     Q_2  = Q_3 \Upsilon_d^* .
\end{eqnarray*}
{}From the above equations after some manipulations we obtain
\begin{eqnarray*}
&(\Upsilon_\nu \Upsilon_\nu^*)   Q_1  =  Q_1 (\Upsilon_u \Upsilon_u^*),
\qquad &
(\Upsilon_e    \Upsilon_e^*)     Q_2  = Q_2 (\Upsilon_d \Upsilon_d^*),\\
&(\Upsilon_\nu^* \Upsilon_\nu)   Q_3  =  Q_3 (\Upsilon_u^* \Upsilon_u),
  \qquad &
   (\Upsilon_e ^*   \Upsilon_e)     Q_3  = Q_3 (\Upsilon_d^* \Upsilon_d)\ .
\end{eqnarray*}

Thus in order that the algebra generated by 
$A_{l g}^{(2)}$ and $D_g = D_{lg} \oplus D_{qg}$
is exactly $M_{4g} \oplus M_{4g}$, 
it suffices then that the only solutions of these equations are 
$Q_1 = Q_2 = Q_3 = 0$.

We shall derive here only the sufficient condition, which we later confront with the physical parameters (measured in experiments).

Due to the diagonal form of the mixing matrices $\Upsilon_e$
and $\Upsilon_u$, and unitarily diagonalizable form of the mixing 
matrices $\Upsilon_\nu$ and $\Upsilon_d$, it is straightforward to 
verify that whenever any eigenvalue 
of $\Upsilon_\nu$ is distinct from any eigenvalue 
of $\Upsilon_u$, and  any eigenvalue 
of $\Upsilon_e$ is distinct from any eigenvalue 
of $\Upsilon_d$, the above equations have only zero solution.

More precisely, if $U$ diagonalizes 
$\Upsilon_\nu = U^* \Upsilon_\nu^{diag} U$, 
then the first identity becomes 
$$({\Upsilon_\nu^{diag}}\Upsilon_\nu^{diag})^* \, UQ_1=UQ_1 (\Upsilon_u \Upsilon_u^*).$$ 
If the eigenvalues of respective diagonal matrices $\Upsilon_\nu^{diag}$
and $\Upsilon_u$ are different from each other then as a consequence 
$Q_1 = 0$ and $Q_3=0$.  Next, observe that a similar argument works 
for the second identity for $\Upsilon_e$ and $\Upsilon_d$, from
which we obtain $Q_2=0$ and $Q_3=0$.

Hence if the matrices $\Upsilon_\nu, \Upsilon_e$ and $\Upsilon_e, \Upsilon_d$
have different eigenvalues (in each pair) then the only solution is $Q_1=Q_2=Q_3=0$
and as a consequence,  the Hodge property is satisfied.  

To summarise, in addition to the conditions in the previous subsection 
on the masses and mixing matrix of leptons and on the masses and mixing 
matrix of quarks, if all the up leptons (neutrinos) masses are different from any of the 
masses of up quarks and similarly for thee masses of down leptons  and of down quarks then the Hodge condition is satisfied.

We note that this condition could be relaxed, it is, in particular not necessary,
that all masses need to be different from each other. However, as our aim is 
to verify whether the physical parameters do lead to the Hodge property we 
omit the detailed discussion of precise necessary and sufficient conditions.

\section{Does Standard Model (with the currently known parameters) 
	satisfy the Hodge duality ?}

We will now analyse the experimental data in Standard Model.

In the physical case when $g=3$, $\Upsilon_e$ is the $e,\mu,\tau$ mass matrix (which following usual conventions we assume to be diagonal) and $\Upsilon_\nu$ is the mass matrix of corresponding  neutrinos, which is twisted by the unitary Pontecorvo--Maki--Nakagawa--Sakata mixing matrix (PMNS matrix),
and $\Upsilon_u$ is the up quarks $u, c, t$ mass matrix (which we assume
as diagonal), and $\Upsilon_d$ is the down quarks 
$d, s, b$ mass matrix, which is twisted by the unitary Cabibbo--Kobayashi--Maskawa mixing matrix (CKM matrix).

\subsection{Lepton sector}

Consider first the leptonic sector. Here, as aforementioned the physicists 
convention is to diagonalize "down" leptons and the PMNS matrix mixes 
the neutrinos. Then $\Upsilon_e=\delta^{lep}_{\downarrow}$ 
and 
$\Upsilon_\nu= U\delta^{lep}_{\uparrow}U^*$, with diagonal non-negative $\delta^{lep}_{\uparrow}$ ($\to$ up lepton masses) and $\delta^{lep}_{\downarrow}$ ($\to$ down lepton masses), and unitary $U$. Using the most recent results
\cite{Cap} and the standard convention to parametrise the the PMNS matrix using
three mixing angles $\theta_{12},\theta_{13},\theta_{23}$ and a Dirac phase $\delta$,
$$ U = \begin{bmatrix} c_{12}c_{13} & s_{12} c_{13} & s_{13}e^{-i\delta_{}} \\
-s_{12}c_{23} - c_{12}s_{23}s_{13}e^{i\delta_{}} & c_{12}c_{23} - s_{12}s_{23}s_{13}e^{i\delta_{}} & s_{23}c_{13}\\
s_{12}s_{23} - c_{12}c_{23}s_{13}e^{i\delta_{}} & -c_{12}s_{23} - s_{12}c_{23}s_{13}e^{i\delta_{}} & c_{23}c_{13} \end{bmatrix}, 
$$
where $s_{ij} = \sin \theta_{ij}$ and $c_{ij} = \cos \theta_{ij}$.

As the parametrisation of the matrix is chosen so that all angles are smaller
than $\frac{\pi}{2}$, then the only possibility that the Hodge duality is broken 
in the leptonic sector is that at least two of the angles vanish. 
However, within the $3\sigma$ range
(using data from \cite{Cap} with normal ordering of masses assumed\footnote{The other possibility, inverted ordering, changes the values of angles by less than 5\%, so it does not change the conclusions.}) we have:
\begin{center}
\begin{tabular}{|c|c|c|}
\hline
$ 10 \sin^2 \theta_{12} $ &  $10^{2} \sin^2 \theta_{13} $& $10 \sin^2 \theta_{23}$ \\ \hline
 2.65 -- 3.46 & 1.90 -- 2.39 &  4.30 -- 6.02 \\ \hline
\end{tabular}
\end{center}
and we see that albeit one of them is very small, they all are still non-zero.

Note that depending whether the massive neutrinos are Dirac or Majorana 
there  may be two additional phases which were not measured so far.
Nevertheless, it is clear that independently of the phases, within the measured 
accuracy, the Hodge condition is not broken. Since the masses of electron, muon and tau are different from each other, then, the lepton mixing is maximal and the Clifford algebra generated in the leptonic sector is the full matrix algebra $M_{4g}$, provided that no neutrino mass vanishes.

Thus vice versa, if the quantum analogue of such a geometric property we named Hodge duality is to be satisfied, the non vanishing of neutrino mass can be regarded as a particular prediction for the Standard Model. Otherwise the multiplet of fundamental fermions won't have a clear geometric status, neither of quantum spinors, nor de\,Rham forms.

Note that the current data on experimental measurement of the so-called Jarlskog invariant, which more conveniently measures the CP-violation,
$$ J_{\nu CP}^{max} = \frac{1}{8} \cos(\theta_{13}) \sin(2\theta_{13}) \sin(2\theta_{23}) \sin(2\theta_{12}),$$ 
provides for the neutrino mixing range (with 1$\sigma$ error):
$$ J_{\nu CP}^{max} = 0.0329 \pm 0.0007 ,$$
that proves indeed that not only all the angles are non-vanishing but all 
neutrino masses are different from each other. 

The absolute values of the neutrino masses can only come from experiments
that provide non-oscillation data (from single $\beta$ decay for example, or
cosmology). Currently there are no conclusive results here both as to the nature
of neutrinos (Majorana or Dirac) or the absolute mass scales (with only upper limits on the mass scale). 
Therefore the Hodge condition, which suggests that all neutrino masses are nonzero should be considered as an engaging prediction.

\subsection{Quark sector}

Here the usual convention in physics is different with "up" sector diagonal 
and "down" sector mixed. The bare up and down quark massed are different 
from each other within the errors so the only thing to check is the mixing 
matrix $U$. Using again the same type of parametrization of the matrix
by three angles and the phases, we can just look at the experimental value
of the Jarlskog invariant $J_{qCP}^{max}$, measured with 1$\sigma$ \cite{Pat},
$$ J_{qCP}^{max} = \left( 3.04 \bfrac{+0.21}{-0.20} \right) \, 10^{-5}, $$
which is sufficient to ascertain that all angles are indeed nonzero and
that implies the partial condition to Hodge duality.

Note that unlike in the leptonic case the angles are very small, which means
that the matrix $U$ is very close to the diagonal unit matrix. Nevertheless 
within the experimental errors we see that the quark mixing is also maximal 
and the Clifford algebra generated in the quark sector is the full matrix 
algebra $M_{4g}$ as well.

\subsection{Full Hodge duality}

Finally we inspect the lepton plus the quark sectors together. Since within the experimental error no lepton mass equals to some quark mass, by the analysis at the end of previous section the algebra generated by $A_{l}^{(2)}$ and $D_{l} \oplus D_{q}$
is exactly $M_{4g} \oplus M_{4g}$ and the Hodge property holds consequently for the entire Standard Model with three families, provided 
that there is no massless neutrino.

Observe that these conditions are sufficient and since they are satisfied for the Standard Model we do not need to analyse and compare the mixing 
of the leptonic and the quarks sector.

\section{Conclusions}

We have established that the Hodge condition which is a quantum analogue of the geometric condition that characterizes de\,Rham differential forms, 
is satisfied by the fundamental Fermions in the Standard Model under the 
proviso that neutrinos are not massless. It will be interesting to establish in the "bottom-up scenario" if the  Hodge property of the experimental values at a 
low energy scale of quark and neutrino masses and mixing parameters are 
preserved under  the renormalization group running (see eg. \cite{OZ14} and references therein for the neutrino mixing and \cite{DS90}).

Vice versa, the Hodge condition enforces the masses to be different and non-zero, and the nontrivial mixing in the quark sector and in the leptonic sector. Though the Hodge condition appears to be purely geometrical it is thus quite surprising 
that it enforces such physical effects like maximal mixing or non-zero masses.  One can therefore interpret it as a significant feature of the model that is confirmed
by current measurements. Of course, future experiments can possibly falsify the claim about the neutrino masses, nevertheless it is a striking feature that the Standard Model uncovers more refined structure than previously anticipated. This can be used as a guideline towards the construction of possible SM extensions and generalizations. 

 \acknowledgments 
 It is a pleasure to thank Francesco D'Andrea for helpful comments and observations and Serguey Petkov for providing useful references.

\end{document}